# Work Function and High-Coverage Adsorption Energy as Hydrogen-Evolution Descriptors on Ag-Au-Pd-Pt Alloys

Zacharias Liasi,[a] Ridha Zerdoumi,[b,c] Felix Thelen,[b] Geovane Arruda de Oliveira,[c] Rico Zehl,[b] Natalia Pukhareva,[b] Leonardo H. Morais,[b] Henrik H. Kristoffersen,[a] Alfred Ludwig,[b, d] Wolfgang Schuhmann,[c] and Jan Rossmeisl*[a]

[a] Center for High-Entropy Alloy Catalysis (CHEAC), Department of Chemistry
University of Copenhagen
Universitetsparken 5, 2100 Copenhagen, Denmark
E-mail: jan.rossmeisl@chem.ku.dk

[b] Materials Discovery and Interfaces, Institute for Materials, Faculty of Mechanical Engineering
Ruhr University Bochum
Universitätsstraße 150, 44801 Bochum, Germany

[c] Analytical Chemistry - Center for Electrochemical Sciences (CES), Faculty of Chemistry and Biochemistry
Ruhr University Bochum
Universitätsstraße 150, 44801 Bochum, Germany

[d] ZGH and RC FEMS,
Ruhr University Bochum
Universitätsstraße 150, 44801 Bochum, Germany

Supporting information for this article is appended to the end of the document.

**Abstract:** Hydrogen-evolution activity is commonly rationalized through hydrogen adsorption energies and the Sabatier principle, yet this descriptor picture becomes ambiguous on multimetallic surfaces, where each composition exposes a distribution of local adsorption environments. Here we investigate whether the bare-surface work function, previously shown to add predictive information for monometallic surfaces, remains an activity descriptor for compositionally complex surfaces. We test this on three combinatorial Ag-Au-Pd-Pt thin-film materials libraries screened for acidic hydrogen evolution by scanning electrochemical cell microscopy. Graph neural networks provide adsorption-energy distributions and work functions for each measured composition. A work-function-only model explained most of the activity variation (mean $R^2_{\log} = 0.903$), as did a coverage-corrected adsorption model (mean $R^2_{\log} = 0.955$), outperforming dilute adsorption (mean $R^2_{\log} = 0.758$). Combining work function and coverage-corrected adsorption yielded the highest fit quality (mean $R^2_{\log} = 0.969$), but only a small gain over coverage-corrected adsorption alone. For these four metals the coverage-corrected adsorption energy and work function follow a similar trend, producing similar activity rankings, hence including both adds little beyond either one individually, although both are strong predictors.

## Introduction

The hydrogen evolution reaction (HER), the cathode reaction of water electrolysis, is a readily deployable route from renewable electricity to chemical fuels.[1] The search for HER catalysts has for two decades been anchored to a single metric, the adsorption energy of surface hydrogen, $\Delta E_{\mathrm{H}*}$.[2–4] Following the Sabatier principle, a good catalyst binds hydrogen neither too weakly nor too strongly, and plotting activity against adsorption energy across materials yields the canonical volcano.[2, 3] A quantity used this way to organize activity trends across materials is called an activity descriptor, and $\Delta E_{\mathrm{H}*}$ is its archetype.

The dilute hydrogen adsorption energy does not, however, carry all the activity-related information. The weak-binding coinage metals fall well below the activity that their adsorption energies predict (Figure 1D), so part of the trend must come from elsewhere.[2, 5] Adsorption energy is also a local quantity. It describes one adsorbate on one site in a chosen structural model. The work function Φ, on the other hand, the energy required to move an electron from the surface to vacuum, is a property of the bare surface (*i.e.* the surface without any adsorbates) as a whole. It fixes the Fermi level relative to vacuum and is tied to the interfacial electrostatic environment, the potential of zero charge, and electron-transfer barriers.[6] Correlations between HER exchange currents and the work function across the transition metals date back more than half a century.[6, 7] Recently, Østergaard, Bagger, and Rossmeisl made the case quantitative for pure metals.[5] Fitting experimental HER activities with hydrogen and hydroxide adsorption energies and work functions, they showed that the work function adds quantifiable predictive power beyond hydrogen adsorption alone, explaining, for example, why adsorption-energy volcano models overestimate the activity of Cu.[5] Patel *et al.* recently reached a compatible conclusion from first-principles kinetics of alkaline HER on eight late transition metals, where a hybrid descriptor combining the hydrogen adsorption energy with the potential of zero charge – the electrochemical counterpart of $\Phi$, differing from the vacuum work function mainly by screening of the interfacial field – raised the coefficient of determination for the simulated exchange current densities from 0.73 and 0.75 for either quantity alone to 0.92, with the same ordering against the experimental activities (0.72 and 0.71 to 0.86).[8] They also found that the hydrogen adsorption energy for the top site correlates at least as well on its own ($R^2$ = 0.96 and 0.90), demonstrating that an electrostatic term is not the only way to recover what the dilute hollow site adsorption energy misses. Figure 1D-F summarizes this pure-metal picture, a volcano in the adsorption energy and a monotonic trend for the work function.

Whether this result extends to compositionally complex solid solutions (CCSS), including high-entropy alloys (HEAs), remains unresolved, yet these materials now dominate catalyst discovery efforts.[9–11] Multinary and HEA surfaces expose many local surface atom arrangements, so each composition carries a distribution of adsorption energies rather than a single value, and a few near-optimal sites can dominate the activity even when they are rare.[9, 12] Composition, crystal structure, and surface segregation have all been shown to tune HER and hydrogen oxidation activity on such surfaces.[13–15] For the work function, there is a first indication that the pure-metal trend has an alloy counterpart: On a Pt-Pd-Ru-Ir-Ag materials library, the potential of zero charge follows the composition-weighted average work function and correlates with acidic HER activity.[16] That observation is, however, a correlation between measured quantities. It does not test whether the work function carries predictive information beyond the adsorption-energy distribution, which itself varies systematically with composition. This leads to the two questions of this work. First, does the work function act as an activity descriptor for HER on alloys? Second, does it add predictive information beyond the adsorption energy distribution?

We attempt to answer those questions for HER on Ag-Au-Pd-Pt composition-spread thin-film materials libraries.[17] Scanning electrochemical cell microscopy (SECCM), in which a small electrolyte-filled pipette forms a local electrochemical cell on the surface, provides activity measurements at hundreds of compositions on each materials library.[18–20] Graph neural networks (GNN) trained on density functional theory (DFT) data supply for every measurement area (MA) the distribution of hydrogen adsorption energies and the mean bare-surface work function at near-DFT accuracy.[21–23] Using a common kinetic rate equation we compare three models, one built on the adsorption-energy distribution, one built on the work function alone, and one built on the combination of the two.

## Results and Discussion

The applied approach integrates combinatorial thin-film synthesis, physicochemical characterization, and high-throughput electrochemical screening within a unified workflow, as illustrated in Figure 1.[17, 20] Thin-film materials libraries, denoted Lib-1, Lib-2, and Lib-3 within this work, with systematic composition gradients were fabricated by combinatorial co-sputtering from four elemental targets of Ag, Au, Pd, Pt onto a single sapphire wafer as a non-rotating substrate (Figure 1A). From this, a spatially resolved compositional spread across 342 MAs of each library was obtained and subsequently analyzed by high-throughput methods. Each MA was characterized by energy-dispersive X-ray spectroscopy (EDX) for elemental composition and by X-ray diffraction (XRD) for phase and structural analysis (Figure 1C). Lib-3 was originally prepared and characterized for a separate study of phase stability in dependence of thermal processing conditions, in which its composition and HER activity data were used to select one specific composition for further investigation.[24] Here EDX and SECCM data from the as-deposited Lib-3 are used together with two additional materials libraries to investigate electrocatalytic descriptor-activity relationships. Electrochemical activity for the HER was assessed at each MA using a long-range SECCM setup operating in hopping mode (Figure 1B).[20] In this configuration, a quasi-reference counter electrode (QRCE) inside an electrolyte-filled pipette is sequentially approached to each MA, where the electrolyte meniscus establishes a confined electrochemical cell at the working electrode (WE) surface. This geometry allows independent, spatially resolved measurements of electrocatalytic activity across the full compositional gradient of the library without cross-contamination between MAs. Further details about the characterization and activity measurement approaches are given in section S1 in the Supporting Information (SI).

The resulting electrochemical dataset was interpreted in conjunction with the theoretical descriptors to establish structure-activity relationships (Figure 1C). As shown in Figures 1D-F, exchange current density was correlated against hydrogen adsorption energy and work function across a reference set of monometallic catalysts, revealing a classical volcano-type dependence on adsorption energy (Figure 1D) and monotonic linear relationship with work function (Figure 1E). The apparent orthogonality in Figure 1F for pure metals motivates testing whether the same holds on alloys.

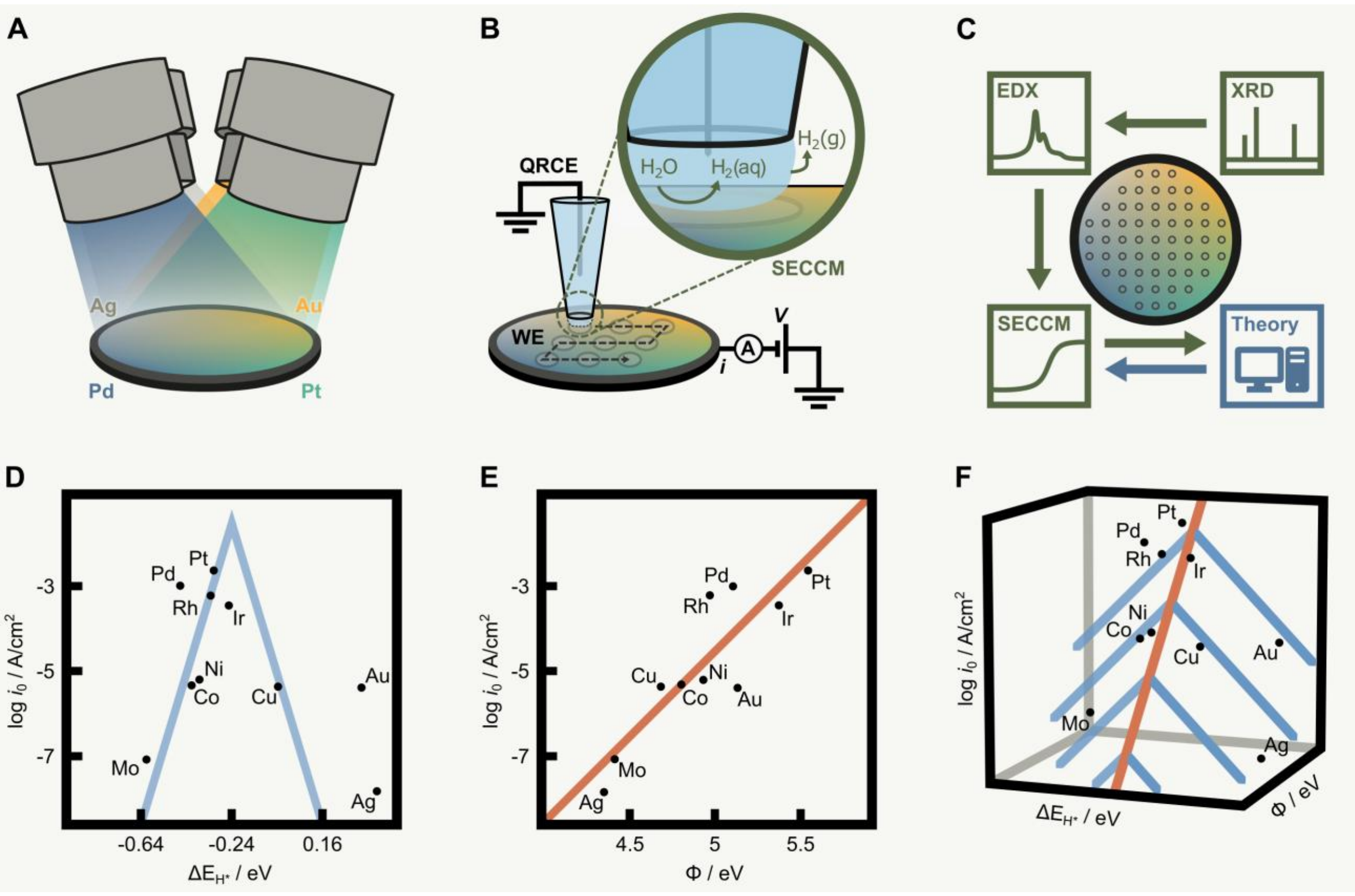

**Figure 1.** Schematic overview of workflow and descriptor motivation. (A) Co-sputter deposition of thin-film materials libraries. Four elemental targets (Ag, Au, Pd, and Pt) are directed simultaneously onto a sapphire wafer, producing a continuous composition gradient across the library. (B) The SECCM principle. The electrolyte-filled pipette forms a confined meniscus cell at each MA of the working electrode (WE), and the potential is applied vs a quasi-reference counter electrode (QRCE) and current is measured using a transimpedance amplifier. The inset shows the HER taking place inside the meniscus. (C) The high-throughput characterization loop. Composition (EDX) and structure (XRD) are recorded for every MA before the SECCM screening, and theory and experiment inform each other through the descriptor models. (D-F) Base descriptor motivation for the adsorption energy and work function. The logarithmic exchange current density, the HER rate at zero overpotential $i_0$, is plotted against the hydrogen adsorption energy (D) and the work function (E), with (F) showing the combined three-dimensional descriptor space. Exchange current densities are taken from Ref. [6], while the adsorption energies and work functions are computed (see the Supporting Information for details). The volcano in (D) and the monotonic trend in (E) motivate testing the two quantities as complementary descriptors. Note that the slopes and volcanoes in D-F are purely illustrative.

### Descriptors

Both descriptors were computed for fcc(111) slabs with a 3×3×4 surface cell. Hydrogen was adsorbed in each of the nine fcc hollow sites in turn, giving nine adsorption energies per slab at 1/9 monolayer (ML) coverage. On pure metals the nine sites are equivalent, so one was used. The electronic adsorption energy per site is

$$\Delta E_{\mathrm{H}^*} = \frac{1}{n}\left(E\left(\mathrm{slab{+}}n\mathrm{H}^*\right) - E(\mathrm{slab}) - \frac{n}{2}E(\mathrm{H}_2)\right),$$

where $E\left(\mathrm{slab} + n\mathrm{H}^*\right)$ is the energy of the slab with $n$ adsorbed hydrogen atoms, $E(\mathrm{slab})$ is the energy of the bare slab, and $E(\mathrm{H}_2)$ is the energy of gas-phase molecular hydrogen at the same level of theory. From the bare slabs, the work function was calculated as

$$\Phi = V_{\mathrm{vac}} - E_{\mathrm{F}}.$$

Here $E_{\mathrm{F}}$ is the Kohn-Sham Fermi level and $V_{\mathrm{vac}}$ is the Hartree electrostatic potential averaged over the surface plane at the cell boundary 10 Å from the surface. A dipole correction was applied along the surface normal. The structures were first relaxed with the UMA-S-1.2 universal machine-learning interatomic potential,[25] and subsequently relaxed and evaluated using the RPBE exchange-correlation functional,[26] as implemented in the grid-based projector-augmented waves (GPAW) software package.[27] Atomic structures were handled using the Atomic Simulation Environment (ASE).[28] The full relaxation settings, cell parameters, cutoffs, and convergence criteria are given in section S2 in the SI.

The composition space was sampled on systematic grids along the binary edges and ternary faces and with a Sobol sequence[29] through the quaternary interior. This gives 582 compositions, represented by 1,738 symmetry-distinct slabs, yielding 1,738 work function values and 15,610 adsorption energies (Section S2). Two graph neural networks, one per descriptor, were trained on the data to predict the relaxed value from the unrelaxed geometry. Both use the DimeNet++ architecture[22, 23] on periodic graphs. Holding out whole compositions, the adsorption-energy model reaches a test mean absolute error of 0.0093 eV ($R^2$ = 0.994) and the work-function model 0.0179 eV ($R^2$ = 0.988), measured against the RPBE values (Section S3). The atomic arrangement of a MA is unknown, so each MA was represented by an ensemble of 1,000 random slabs drawn from its EDX composition, giving 9,000 adsorption energies and 1,000 work functions per MA (Section S4). All four layers are drawn from the same distribution, so surface segregation is not modeled. Both descriptors come from the same slabs, so this bias largely cancels in the comparison between them. These are 1/9 ML energies and therefore miss the H-H repulsion expected at the higher local coverages present during the measurement.[8, 30, 31] We convert them to 1 ML values with a Frumkin-type mean-field correction, in which lateral repulsion weakens adsorption as the surface fills:

$$\Delta E_{\mathrm{H}^*,p,q}(\theta') = \Delta E_{\mathrm{H}^*,p,q}(\theta) + \omega(\theta' - \theta).$$

Here $\theta$ is the dilute fractional coverage of the inferred adsorption energy and $\theta'$ is the near-saturated target coverage. The lateral H-H interaction coefficient $\omega$ = 0.25 eV/ML was obtained by filling equimolar Ag-Au-Pd-Pt(111) slabs with hydrogen one atom at a time (Section S2). It equals the mean of the four pure metals to within 0.003 eV/ML, and our Pt value (0.183 eV/ML) agrees with the parameter used in recent first-principles kinetics of hydrogen evolution (0.188 eV/ML, calculated for Pt(111)).[8, 30] For the conversion from 1/9 ML to 1 ML used in the main model, this corresponds to a rigid shift of 0.22 eV, which displaces the adsorption-energy distribution uniformly toward weaker binding without altering its shape.

**Activity Models**

We employ the canonical adsorption energy optimum $\Delta E_{\mathrm{H}*,\mathrm{opt}} = -0.24\,\mathrm{eV}$, corresponding to $\Delta G_{\mathrm{H}*} = 0\,\mathrm{eV}$ using the usual zero-point-energy and entropy correction of 0.24 eV, throughout this work.[2] For adsorption site $q$ in composition $p$, the coverage-corrected adsorption energy is scored based on its absolute distance from the electronic adsorption energy optimum through the exponential function

$$f_{p,q} = \exp\left(\frac{-\alpha_{\mathrm{ads}}\left|\Delta E_{\mathrm{H}*,p,q}(\theta') - \Delta E_{\mathrm{H}*,\mathrm{opt}}\right|}{k_{\mathrm{B}}T}\right).$$

Here $k_{\mathrm{B}}$ is the Boltzmann constant, $T$ is temperature in Kelvin, and the product is 0.0257 eV at 298.15 K. The function produces a symmetrical exponential volcano, with $\alpha_{\mathrm{ads}}$ controlling the width, $\Delta E_{\mathrm{H}*,\mathrm{opt}}$ controls the placement of the volcano center, and the exponential function ensures that even small differences in adsorption energy of near optimal sites are distinguishable. $\Delta E_{\mathrm{H}*,p,q}(\theta')$ is the hydrogen adsorption energy of site $q$ in composition $p$ at fractional coverage $\theta$'. Averaging over all sites used to represent a given composition $p$ yields the adsorption-energy-based descriptor term for that composition

$$f_p = \frac{1}{N_p}\sum_{q=1}^{N_p} f_{p,q},$$

with $N_p$being the total number of sites. The average is taken after applying the exponential site factor. This is important because the volcano is nonlinear. A small number of sites close to the optimum can contribute more strongly than the mean adsorption energy alone would suggest.[9]

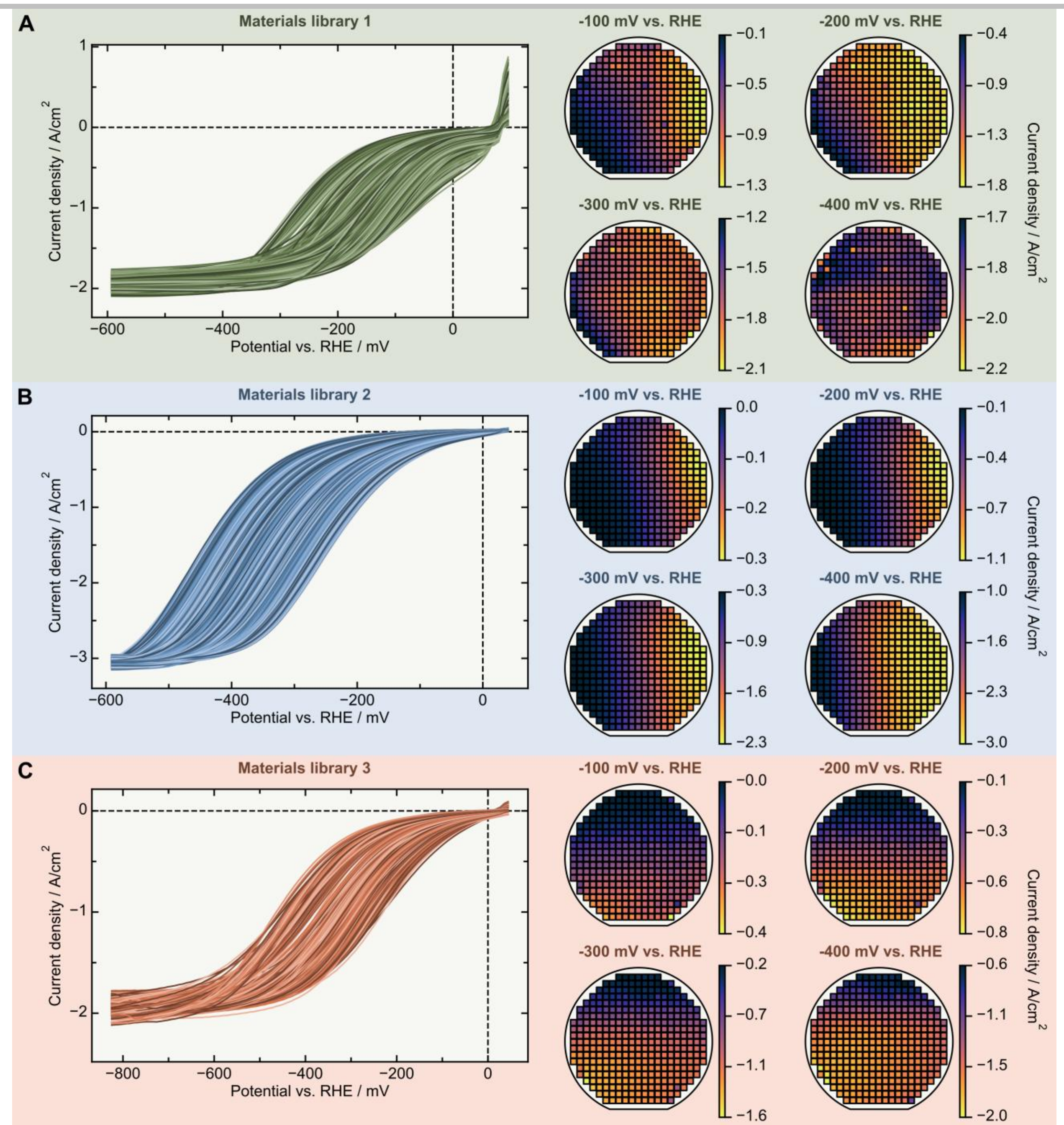

**Figure 2.** SECCM screening of the three materials libraries. Left, all linear sweep voltammograms for (A) Lib-1, (B) Lib-2, and (C) Lib-3. Right, materials library maps of the current density at -100, -200, -300, and -400 mV versus RHE. Each library contains 342 MAs, measured in 0.1 M $HClO_4$ with 0.1 M $LiClO_4$, the current densities are normalized by the pipette opening area.

A similar function is used to score compositions based on their work function. The work function is a property of the bare surface as a whole, allowing us to score the compositions directly, as

$$g_p = \exp\left(\frac{\alpha_{\mathrm{wf}}\left(\Phi_p - \Phi_{\mathrm{ref}}\right)}{k_{\mathrm{B}}T}\right).$$

The fitted parameter $\alpha_{\mathrm{wf}}$ is dimensionless and controls the rate at which the predicted activity changes with work function. A positive value means that compositions with a higher work function are more active. $\Phi_p$ is the work function of composition $p$, inferred as an ensemble average across a finite set of slabs. $\Phi_{\mathrm{ref}}$ is an arbitrary reference value setting the point at which $g_p$ equals unity. Here the work function of Pt(111) was used $\Phi_{\mathrm{ref}} = 5.50$ eV, so that $j_0$ is the exchange current density of a surface with that work function.

Using either or both descriptor functions, current density values are obtained by introducing a fitted reference exchange current $j_0$ with the unit $\mathrm{A/cm^2}$, effectively a scaling factor, and a Tafel-style applied-potential dependence through

$$h(\eta) = \exp\left(\frac{-\alpha_c\, e\eta}{k_B T}\right).$$

Here $\alpha_c$ is the cathodic charge transfer coefficient, $e$ is the elementary charge, and $\eta$ is the overpotential. The kinetic current density, as predicted by hydrogen adsorption energy alone, for a given alloy composition is thus

$$j_{\mathrm{kin},p}^{(\mathrm{A})}(\eta) = -j_0\, h(\eta) f_p,$$

denoted model A. Likewise, model B, using only the work function as a descriptor of activity is written as

$$j_{\mathrm{kin},p}^{(\mathrm{B})}(\eta) = -j_0\, h(\eta) g_p,$$

and the combination of the two yields model C, given by

$$j_{\mathrm{kin},p}^{(\mathrm{C})}(\eta) = -j_0\, h(\eta) f_p\, g_p.$$

At high cathodic current, the measured SECCM curves approach a limiting current (Figure 2). This is included in the modeling through a Koutecký-Levich-type expression,[32]

$$j_p(\eta) = -\frac{\left|j_{\mathrm{kin},p}(\eta)\right|}{1 + \frac{\left|j_{\mathrm{kin},p}(\eta)\right|}{j_{\mathrm{lim}}}}.$$

When the kinetic current is small compared with $j_{\mathrm{lim}}$ the denominator is close to one and the current follows the kinetics. When the kinetic current is large the expression saturates at $-j_{\mathrm{lim}}$, the transport limited plateau. We treat it as an effective ceiling for each composition or materials library rather than tying it to a single transport mechanism. The limiting current is the last fitted quantity.

**Activity Predictions**

The three models were fitted jointly to the SECCM current densities extracted from the linear sweep voltammograms at $\eta = -100, -200, -300,$ and -400 mV. The fits were carried out in log-current space, minimizing a Huber loss over all MA-potential pairs, with a common current floor and valid-cell mask across all models.[33] Fit quality is reported as $R^2_{\log}$, the coefficient of determination in $\log_{10}$-current space. Details of the fitting procedure are given in Section S5 in the SI. Table 1 summarizes the three main models at both the dilute (1/9 ML) and coverage-corrected (1 ML) adsorption energies, the latter shifted by 0.22 eV, and Figure 4 shows the corresponding parity plots. All three models reproduce the within-library activity to high fidelity. The work-function model B alone reaches a mean $R^2_{\log}$ of 0.903 (Table 1), so on these multi-metallic surfaces the bare-surface work function is by itself a strong activity descriptor. The coverage-corrected adsorption model A is stronger still (mean 0.955), and the combined model C is the strongest (mean 0.969). The fitted volcano width of model A is a near or at unity slope ($\alpha_{\mathrm{ads}}$ = 1, 0.845, and 0.712 across the libraries), with the coverage correction pushing it towards ideal $k_B T$ width, and the work-function coupling in model C, including coverage-correction, is relatively small ($\alpha_{\mathrm{wf}}$ = -0.003, 0.075, and 0.061). The fitted cathodic charge-transfer coefficients fall between 0.31 and 0.46 across models and libraries (Figure 4), corresponding to Tafel slopes of 129 to 188 mV/dec. These are large for acidic HER on platinum-group surfaces. Comparable values arise in first-principles microkinetic models once the coverage dependence of the electrochemical adsorption barrier is included, because the barrier stiffens as coverage rises with potential and damps the potential response of the rate.[8] The magnitude of the fitted slopes is therefore consistent with the coverage effect that the adsorption correction describes.

**Table 1.** Within-library fit quality ($R^2_{\log}$) of the adsorption-only (A), work-function-only (B), and combined (C) models for the three libraries and their mean.

| Model | Coverage | Lib-1 | Lib-2 | Lib-3 | Mean |
|---|---|---|---|---|---|
| A | 1/9 ML | 0.712 | 0.751 | 0.810 | 0.758 |
| A | 1 ML | 0.976 | 0.932 | 0.958 | 0.955 |
| B | -- | 0.909 | 0.888 | 0.912 | 0.903 |
| C | 1/9 ML | 0.938 | 0.896 | 0.932 | 0.922 |
| C | 1 ML | 0.976 | 0.963 | 0.969 | 0.969 |

The interplay between the two descriptors is governed by where the adsorption volcano is centered (Table 1 and Figure 3). Using dilute adsorption energy distributions, model A (adsorption energy alone) is only moderately predictive. Every MA contains some amount of Pt and Pd, and the relevant part of the adsorption energy distributions therefore cluster too closely to be resolved from one another in the dilute regime. At higher coverage (1 ML), Pt and Pd sites -- even if only a few are present -- carry most of the MA's activity representation in the simulated composition space. For the dilute regime, adding the work function through the combined model C lifts the fit to a mean 0.922, a gain of 0.164 over adsorption energy alone. The fitted coupling $\alpha_{\mathrm{wf}}$ is positive and sizable across all three libraries (0.265, 0.146, 0.131), *i.e.*, compositions with a higher work function are predicted to be more active, in line with the pure-metal trend of Figure 1E. The improvement of Model C over the work-function-only model B is by contrast small (0.029, 0.008, 0.020). With the dilute adsorption energies the bare work function is the stronger single descriptor, with a mean $R^2_{\log}$ of 0.903, while the adsorption term contributes little until coverage is corrected to better match the coverage under experimental conditions. Correcting the inferred 1/9 ML adsorption energies for the finite hydrogen coverage present under operating cathodic bias, a rigid weakening of the adsorption energy toward the 1 ML limit, absorbs most of this work-function advantage. With the coverage-corrected adsorption energy distributions, shifted by 0.22 eV, the adsorption-only model A jumps to 0.976, 0.932, 0.958 (mean 0.955), recovering on its own all and more of what the work function had supplied to the dilute model. The combined model C then improves over corrected adsorption energy alone only marginally (0, 0.031, 0.011), and its fitted work-function coupling collapses toward zero ($\alpha_{\mathrm{wf}} = -0.003,\ 0.075,\ 0.061$). In other words, the predictive content that the work function adds beyond the dilute adsorption energy overlaps with the effect of hydrogen coverage. Fitting the adsorption energy distribution shift confirms this picture and identifies the coverage the data prefer within our coverage-correction approximation. For the combined model, the fitted shift converges to 0.22, 0.20, and 0.19 eV across the three libraries. Assigning these shifts entirely to coverage through the mean-field interaction coefficient ($\omega = 0.25$ eV/ML) places the effective working coverage at roughly 0.9 ML, nearly coinciding with the fixed 1 ML ansatz. The fitted-coverage-correction version of model C gives the best fit of all, $R^2_{\log} = 0.976,\ 0.963,\ 0.980$ (mean 0.973), and is the strongest model on every library.

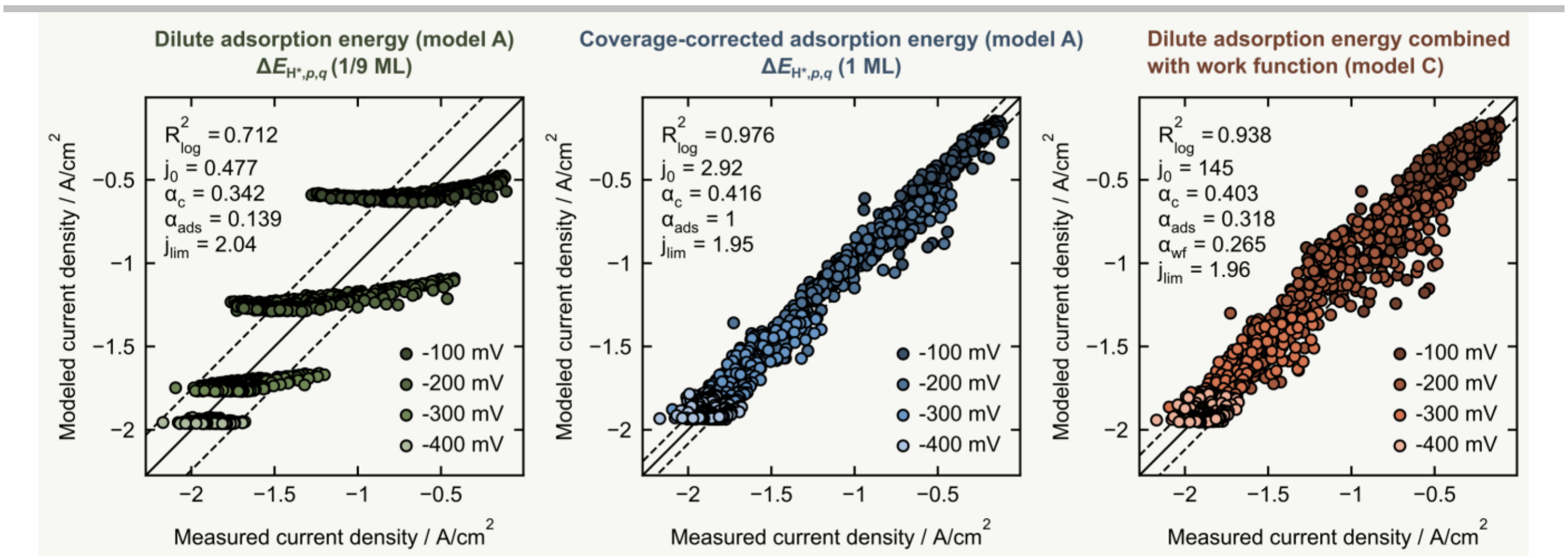

**Figure 3.** Model A versus SECCM current density using non-corrected (1/9 ML) and coverage-corrected adsorption energy distributions (1 ML, shifted by 0.22 eV), and model C using the non-corrected adsorption energies, all Lib-1.

**Transferability Across Libraries**

To test whether the descriptor-based activity model is a genuine, transferable relationship rather than a per-library fit, we performed leave-one-library-out (LOLO) cross-fitting (Figure 5). The descriptor-dependent physics parameters were fitted on two of the three libraries and applied to the held-out third, at three levels of adaptation, no recalibration, recalibration of the per-library pseudo-exchange current $j_0$ only, and recalibration of both $j_0$ and the transport ceiling $j_{\mathrm{lim}}$. Without recalibration the transfer is poor and, on average, negative for every model, because $j_0$ absorbs library-specific offsets in active-site density and cell coupling that do not transfer. Recalibrating only these two scalars restores essentially the full within-library quality, so the shape of the activity prediction itself transfers. For the main coverage-corrected models the transfer gaps are 0.007 for model A ($R^2_{\mathrm{log}}(j_0, j_{\mathrm{lim}})$ = 0.948), 0.024 for model B ($R^2_{\mathrm{log}}(j_0, j_{\mathrm{lim}})$ = 0.879), and 0.017 for model C ($R^2_{\mathrm{log}}(j_0, j_{\mathrm{lim}})$ = 0.953), as listed in Table 2. The combined model with a fitted coverage-correction transfers best of all, reaching $R^2_{\mathrm{log}}(j_0, j_{\mathrm{lim}})$ = 0.954 (0.969, 0.938, and 0.955 for the held-out libraries, a gap of 0.019 below the within-library fits), while the fitted coverage-correction adsorption-only model reaches $R^2_{\mathrm{log}}(j_0, j_{\mathrm{lim}})$ = 0.947. The spread across resampled cross-fits was negligible (standard deviation of $R^2_{\mathrm{log}}$ at most 0.001 throughout). Full LOLO details are provided in Section S5 in the SI.

**Table 2.** Mean $R^2_{\mathrm{log}}$ values for the different LOLO transfer levels (no recalibration, recalibration of $j_0$ only, and recalibration of both $j_0$ and $j_{\mathrm{lim}}$.) across the three models, using coverage-corrected adsorption energies.

| | No recalibration | $j_0$ recalibrated | $j_0$ and $j_{\mathrm{lim}}$ recalibrated |
|---|---|---|---|
| Model A | -0.569 | 0.882 | 0.948 |
| Model B | -0.358 | 0.822 | 0.879 |
| Model C | -0.405 | 0.893 | 0.953 |

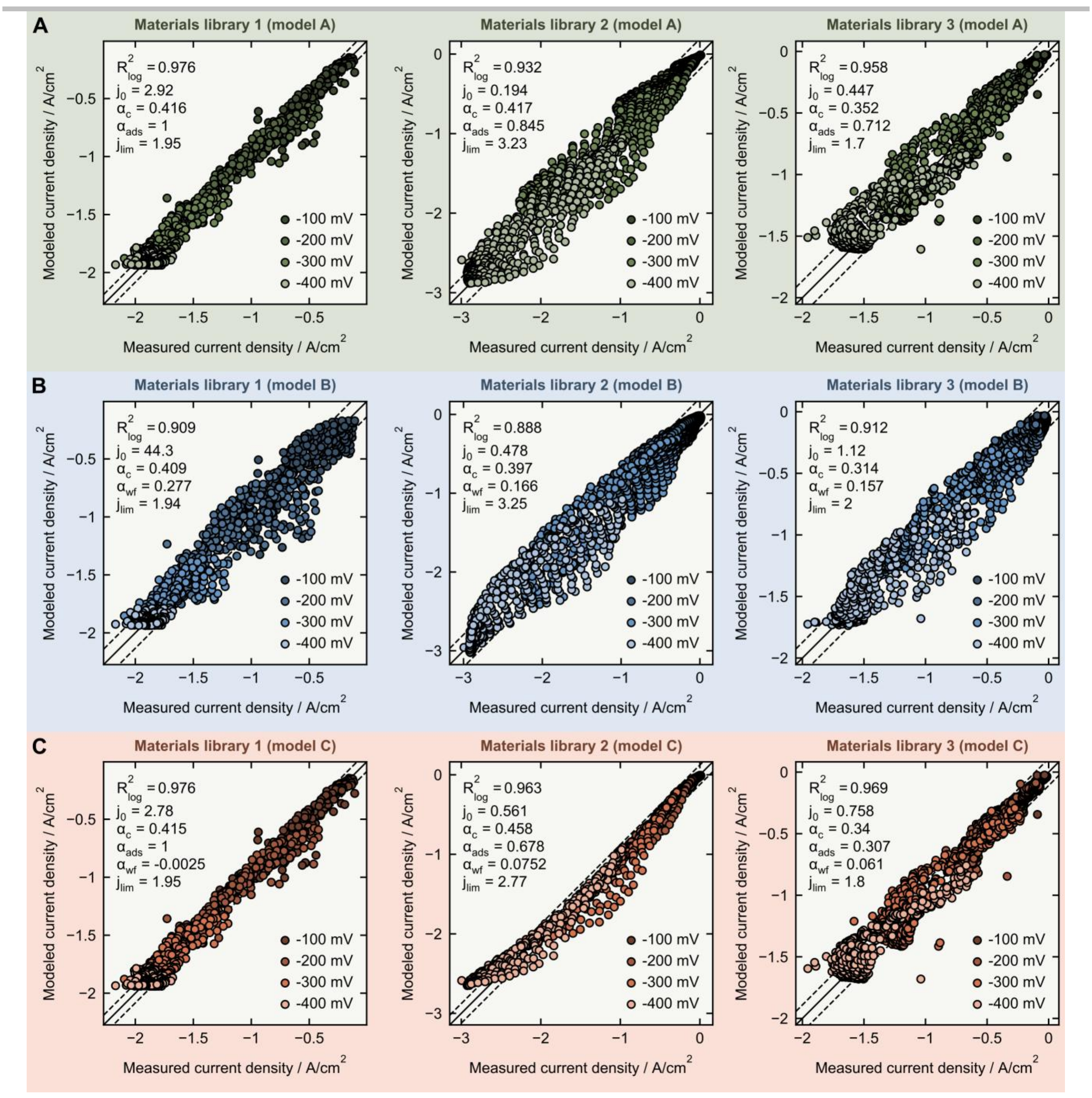

**Figure 4.** Predicted activity fitted to SECCM LSV data for each of the three materials libraries. (A) Model A, (B) model B, and (C) model C versus SECCM current density, using the coverage-corrected energy distributions, shifted by 0.22 eV.

### Work Function as an Independent Descriptor

The gain from the work function could in principle arise trivially, if it acted merely as an extra adjustable degree of freedom. We rule this out with a permutation control in which the per-MA work functions were shuffled across compositions within each library, preserving the library's work-function distribution but destroying its MA-wise pairing with the measured activity, and the combined model was refit on the shuffled data (Table 3). Against the dilute adsorption term, the work function carries decisive and genuinely MA-resolved information. The true combined model ($R^2_{\log}(j_0, j_{\lim})$ = 0.892) far exceeds the adsorption-only reference (0.752), shuffling collapses it back to that reference, and none of 600 within-library or 300 cross-fit shuffles matched the true fit. The work-function contribution is therefore not a free-parameter artefact. It is a real property of the surface, and the pure-metal correlation between work function and HER activity extends to the alloys. Its independence from the adsorption energy is, however, limited. Against the coverage-corrected adsorption term, shuffling no longer degrades the fit to the same extend. The shuffled mean falls back onto the adsorption-only reference, as it does at 1/9 ML, but here the true model C exceeds that reference by only 0.014 within library and 0.005 under transfer.

**Table 3.** Work-function shuffle control. Model C refitted with the per-MA work functions shuffled across compositions within each library, at both coverages. The presented values are mean $R^2_{\log}$. The parentheses in the last column denote the number of shuffled runs reaching or exceeding the unshuffled model C per library.

| Coverage | Fit type | Model A | Model C | Model C, shuffled $\Phi$ |
|---|---|---|---|---|
| 1/9 ML | Within-library | 0.758 | 0.922 | 0.759 (0/600) |
| 1/9 ML | LOLO with both $j_0$ and $j_{\mathrm{lim}}$ refit | 0.752 | 0.892 | 0.752 (0/300) |
| 1 ML | Within-library | 0.955 | 0.969 | 0.955 (161/600) |
| 1 ML | LOLO with both $j_0$ and $j_{\mathrm{lim}}$ refit | 0.948 | 0.953 | 0.948 (200/300) |

**Descriptor Collinearity and Work Function Signal**

The shuffle control shows that the combined-over-adsorption gain vanishes once the adsorption energy is coverage-corrected, but does not explain why. Two further controls make the mechanism explicit. First, the descriptors are collinear on these alloys. Across each library the per-MA mean adsorption energy and the work function are strongly anticorrelated, with $R^2$ = 0.80, 0.73, and 0.51 for Lib-1, Lib-2, and Lib-3 (Pearson correlation of $r$ = −0.90, −0.86, and −0.71). Pooling all 1,025 MAs gives $R^2$ = 0.64, below two of the three libraries, indicating that the relation is curved across the simplex rather than a single global line. Within any one library the two descriptors are therefore less independent than the pooled value suggests. This contrasts with the moderate orthogonality reported for pure metals, where the fcc hydrogen adsorption energy and the potential of zero charge give $R^2$ = 0.26.[8] Here the two descriptors are far from independent. Second, the adsorption-energy-orthogonal component of the work function does not transfer. We residualized the work function against the activity-blind adsorption term, and separately against the composition (EDX) basis, then refit the combined model and scored it within library and under LOLO transfer (Table 4). The component of the work function orthogonal to adsorption does not transfer at any adsorption specification. Its LOLO gain is at most 0.002 and negative at 1 ML. Whatever within-library gain it retains is largest at 1/9 ML (0.065), where it compensates for the uncorrected coverage, but that gain is a per-library adjustment that does not generalize. Once the composition-predictable part of the work function is removed, it adds essentially nothing (0.002 within library and nothing under transfer), so the work function is almost entirely a function of composition. Together with the raw collinearity, this indicates that the standalone predictive power of the work function is carried by its correlation with adsorption energy and composition, not by an independent electrostatic axis.

**Table 4.** Orthogonalized work-function controls. Difference in $R^2_{\log}$ of model C versus model A using the raw work function (model B), the work function residualized against the adsorption energy term $f_p$, and against composition, at 1/9 and 1 ML hydrogen coverage.

| Coverage | $\Phi$ (Direct fit / LOLO with $j_0$ and $j_{\text{lim}}$ refit) | $\Phi \perp \log_{10} f_p$ (Direct fit / LOLO with $j_0$ and $j_{\text{lim}}$ refit) | $\Phi \perp \text{Composition}$ (Direct fit / LOLO with $j_0$ and $j_{\text{lim}}$ refit) |
|---|---|---|---|
| 1/9 ML | 0.164 / 0.140 | 0.065 / 0.002 | 0.003 / -0.055 |
| 1 ML | 0.014 / 0.005 | 0.010 / -0.004 | 0.002 / -0.001 |

**Relation to Pure Metal Studies**

Our starting point was the monometallic result of Østergaard *et al*., a two-dimensional fit of experimental HER activities with the hydrogen adsorption energy and the work function.[5] Both the present work and the recent first-principles study of Patel *et al*. confirm its central claim, that adsorption plus work function outperforms either descriptor alone.[8] Patel *et al*. compiled constant-potential microkinetic models for alkaline HER on eight late transition metals and evaluated descriptors against computed and experimental exchange current densities. They found the potential of zero charge to be a legitimate descriptor on its own ($R^2$ of 0.75 against theory and 0.71 against experiment), a hybrid of fcc adsorption energy and potential of zero charge to be much stronger (0.92 and 0.86), and top-site hydrogen adsorption energy to be their best single descriptor (0.96 and 0.90). They also identified hydrogen coverage and H-H repulsion as real ingredients of the kinetics, adopting a linear repulsion parameter of 0.188 eV/ML calculated for Pt(111),[30] close to our Pt value of 0.183 eV/ML. Their analysis addresses alkaline HER while our measurements are acidic, so the comparison is at the level of descriptors rather than absolute rates.

## Conclusion

For mono-metallic surfaces, the work function adds HER activity information beyond the dilute hydrogen adsorption energy. We investigated whether the same holds for Ag-Au-Pd-Pt alloys, where each composition presents a distribution of adsorption sites rather than a single value, and whether the work function is then more than a free fitting parameter. Using a common kinetic rate equation fitted to SECCM activity across three combinatorial thin-film libraries, the combined descriptor model reproduces every library at least as well as either descriptor alone and best overall ($R^2_{\log}$ up to 0.969) and, trained on two libraries, predicts the held-out third once only the library-specific current scale and transport ceiling are recalibrated ($R^2_{\log}$ = 0.953). Shuffling the work functions between compositions removes the improvement, showing that the work function is a real, spatially resolved surface property carrying activity information and not merely an adjustable parameter. The monometallic result of Østergaard *et al.* thus extends to alloys.

That apparent gain does not, however, survive the coverage correction. The work function and the adsorption energy are correlated across the composition space ($r$ = -0.80, $R^2$ = 0.64, versus the near-orthogonal $R^2$ = 0.26 reported for pure metals). Once the adsorption energies are corrected for the finite working coverage, which the data independently place near 0.9 ML, the combined model improves on adsorption energy alone only marginally, the component of the work function orthogonal to adsorption energy fails to transfer between libraries. On these compositionally complex surfaces the work function is therefore a genuine descriptor of activity, that alone accounts for the majority of the activity when fitted to SECCM data, although the information it provides overlaps with the coverage-corrected adsorption energy. This is the alloy, high-coverage counterpart of a confounding that first-principles studies of pure metals have flagged but could not isolate.

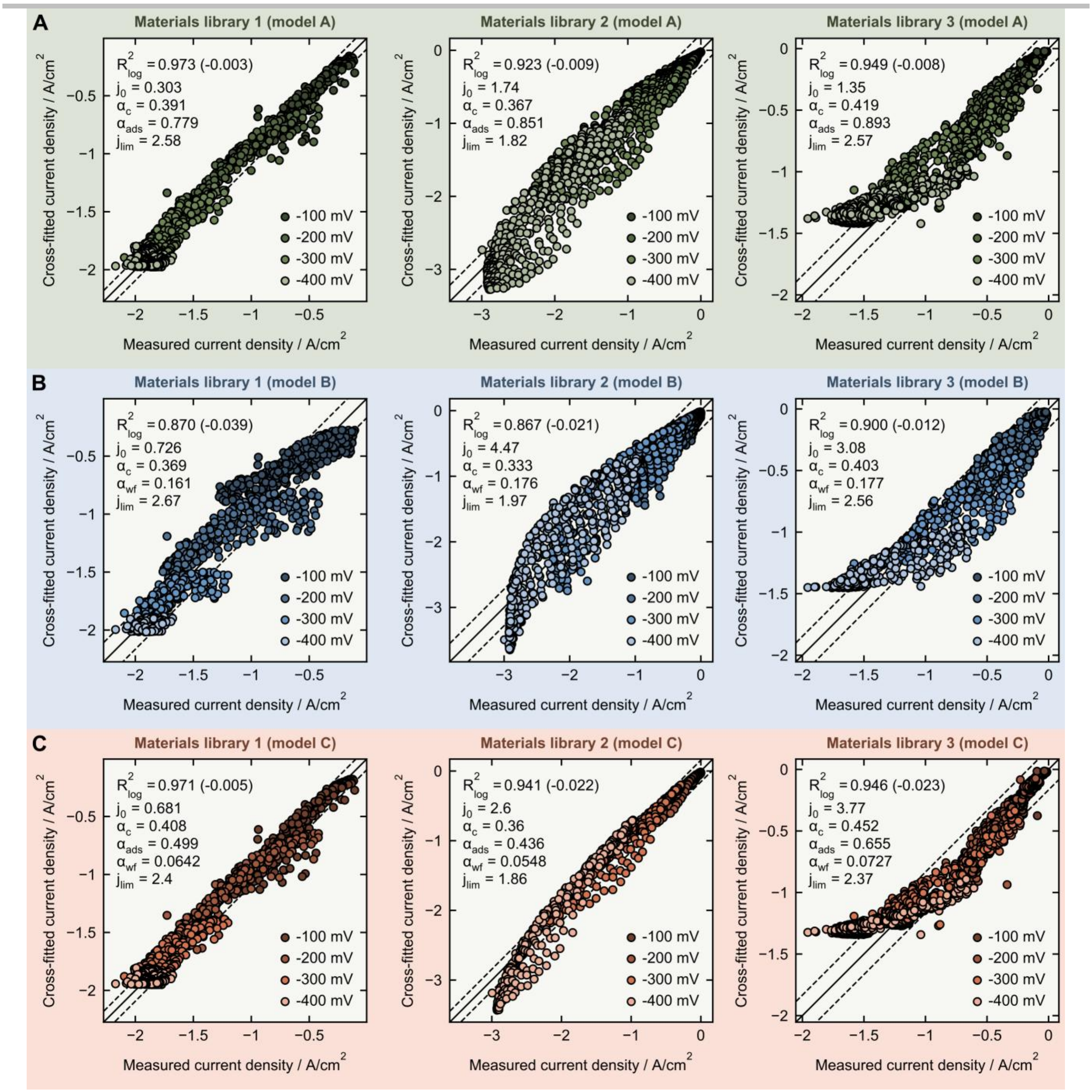

**Figure 5.** Leave-one-library-out fits. Modeled versus SECCM current density as predicted by (A) model A, (B) model B, and (C) model C, using the coverage-corrected adsorption distributions. Each fit is done using parameter values fitted on the two other libraries, with $j_0$ and $j_{\text{lim}}$ recalibrated on the left-out library. The difference between the $R^2_{\log}$ of the left-out library fit and the within-library fit (Figure 4) is given in parentheses.

Thus, both the work function and the coverage correction improve the activity prediction in this alloy space, and to a similar degree. The simplest reading is that neither is the true descriptor of activity, what the models are pointing to is the operating hydrogen adsorption energy, that is, the adsorption energy at the site and coverage the surface holds under reaction conditions. The coverage correction is a direct path towards it, while the work function reaches it through correlation with composition and adsorption energy, which happens to be tight for the selected set of metals. The findings of Patel *et al.* point in the same direction. They found the top site hydrogen adsorption energy to be the best descriptor of activity, outperforming their combined models. In both cases, specifying the adsorption energy the measurements actually sample, be it site or coverage, leaves little to nothing for the electrostatic descriptor to add. This is testable, using an alloy space in which the two quantities, work function and adsorption energy, do not rank the metals the same. For example, an alloy containing Ir and Rh: Ir has the higher work function and Rh the higher exchange current (see Figure 1E), the monotonic trend of the work function would not hold, making it a poor predictor of activity. The usefulness of the work function as a descriptor is likely a property of the metals and not the reaction. Additionally, we have treated the work function as a property of the bare surface as a whole, as it is for pure metals. Whether that holds on compositionally complex surfaces is not known, it may be localized to some degree, making a site-resolved work function informative. More insights into the generalized performance of work function and coverage-corrected adsorption energy across compositionally complex systems are thus needed, as well as what mechanisms or quantities each describe.

## Supporting Information

Experimental details for the thin-film synthesis, the composition and structure characterization, and the SECCM activity measurements (Section S1). The composition sampling, slab set-up, relaxation settings, and the coverage-correction procedure (Section S2). The graph neural network surrogate models together with their graph construction, training protocol, data split, test accuracy, and the composition-only linear baseline (Section S3). The descriptor inference and the ensemble sizes (Section S4). Complete fitting protocol including the loss, valid-cell mask, parameter bounds, and cross-fitting procedure (Section S5). The authors have cited additional references within the Supporting Information.

## Acknowledgements

The authors acknowledge support from the Danish National Research Foundation Center of Excellence, the Center for High-Entropy Alloy Catalysis (Project DNRF149); the Deutsche Forschungsgemeinschaft (DFG, German Research Foundation) through SFB 1625, project number 506711657, subprojects A01, A02, and C01; and the European Union through the European Research Council (ERC) Synergy Grant project 101118768 (DEMI). Views and opinions expressed are, however, those of the authors only and do not necessarily reflect those of the European Union or the European Research Council Executive Agency. Neither the European Union nor the granting authority can be held responsible for them. The authors acknowledge the use of the infrastructure at ZGH (Zentrum für Grenzflächendominierte Höchstleistungswerkstoffe, Center for Interface-Dominated High-Performance Materials) for SEM and XRD measurements.

**Keywords:** activity descriptor • electrochemistry • heterogeneous catalysis • hydrogen evolution reaction • SECCM

## Data Availability

The data that support the findings of this study are openly available in Zenodo at https://doi.org/10.5281/zenodo.22031607.

PREPRINT

# Supporting Information:

# Work Function and High-Coverage Adsorption Energy as Hydrogen-Evolution Descriptors on Ag-Au-Pd-Pt Alloys

Zacharias Liasi,[a] Ridha Zerdoumi,[b,c] Felix Thelen,[b] Geovane Arruda de Oliveira,[c] Rico Zehl,[b] Natalia Pukhareva,[b] Leonardo H. Morais,[b] Henrik H. Kristoffersen,[a] Alfred Ludwig,[b, d] Wolfgang Schuhmann,[c] and Jan Rossmeisl*[a]

[a] Center for High-Entropy Alloy Catalysis (CHEAC), Department of Chemistry
University of Copenhagen
Universitetsparken 5, 2100 Copenhagen, Denmark
E-mail: jan.rossmeisl@chem.ku.dk

[b] Materials Discovery and Interfaces, Institute for Materials, Faculty of Mechanical Engineering
Ruhr University Bochum
Universitätsstraße 150, 44801 Bochum, Germany

[c] Analytical Chemistry - Center for Electrochemical Sciences (CES), Faculty of Chemistry and Biochemistry
Ruhr University Bochum
Universitätsstraße 150, 44801 Bochum, Germany

[d] ZGH and RC FEMS, Ruhr University Bochum
Universitätsstraße 150, 44801 Bochum, Germany

## Contents

## S1 Experimental Methods

### Synthesis and characterization of thin-film materials libraries

The thin-film materials libraries were fabricated at room temperature by magnetron co-sputtering in a combinatorial co-sputter system. Four 2-inch diameter confocal sputter cathodes (AJA International) were equipped with Ag (99.99%, Testbourne), Au (99.99%, Testbourne), Pd (99.95%, Testbourne), and Pt (99.95%, Sindlhauser Materials) targets. Single-side-polished 10 cm diameter c-plane sapphire wafers (SITUS Technicals) were used as substrates. A 15 nm Ta adhesion layer was deposited from a centered sputter cathode. Four DC power supplies (three DCXS-750, AJA International, and one IPP2000, MELEC) were used in power-control mode. The Ar flow during the deposition was 80 sccm at 0.4 Pa. The deposition powers were chosen from preliminary deposition-rate calibrations. The three libraries are labeled Lib-1, Lib-2, and Lib-3. Each library was characterized on 342 predefined measurement areas arranged on an orthogonal array with 4.5 mm spacing. The elemental compositions were measured by energy-dispersive X-ray spectroscopy (EDX) in a JEOL 5800 scanning electron microscope at 20 kV using an Oxford Instruments INCA X-act detector. The acquisition time was 60 seconds per measurement area, at 600 times magnification and a 10 mm working distance. The resulting compositions for all three libraries are shown in Figure S1. X-ray diffraction (XRD) patterns were collected with a Bruker D8 Discover instrument in Bragg–Brentano geometry using a VANTEC-500 area detector at a sample-to-detector distance of 14.9 mm.

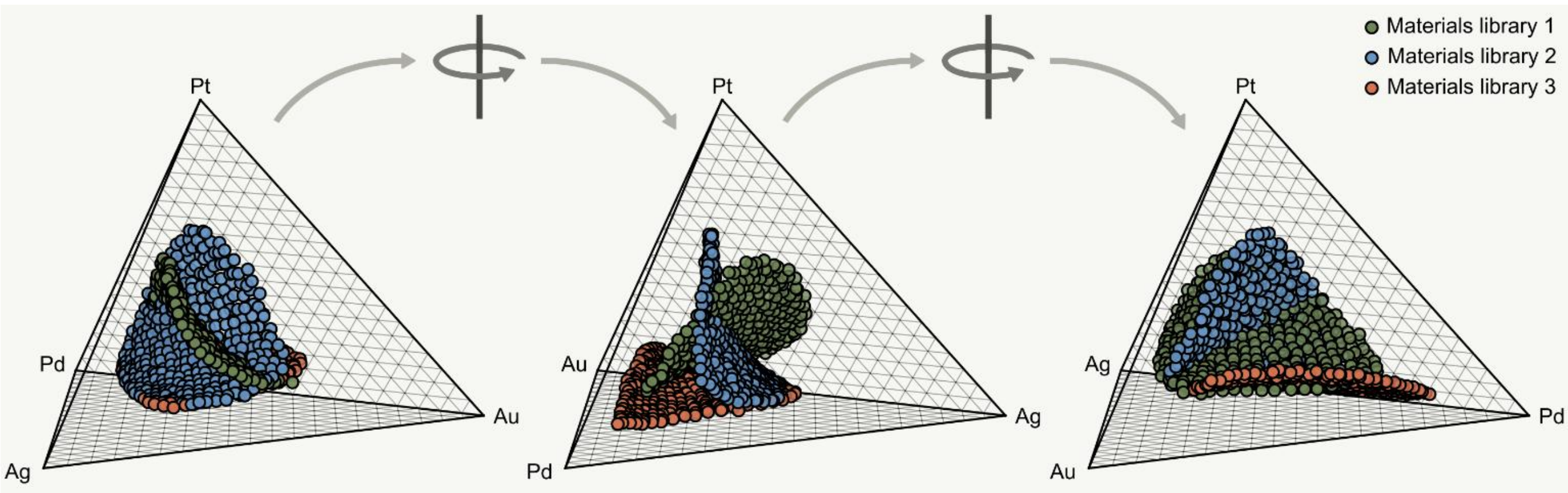

**Figure S1.** EDX measured compositions for each of the three materials libraries considered in this work, seen from three different angles.

### SECCM Measurements

HER activity was measured with an in-house-built long-range SECCM setup operating in hopping mode.[1] Single-barrel quartz filamented capillaries (ZQF-120-90-10, Science Products) were pulled with a $CO_2$ laser puller (P-2000, Sutter Instruments), and the pipette dimensions were checked by scanning electron microscopy (FEI, Quanta 3D ESEM). The electrolyte-filled pipette was mounted on an x, y, z piezo nanopositioning stage (Physik Instrumente), while the materials library, serving as the working electrode, was mounted on a three-axis stepper-motor stage. Three edge positions were used to correct sample tilt during long-range movement between measurement areas. At each measurement area, the SECCM probe was approached until a surface-current threshold of about 10 pA indicated meniscus contact. The meniscus formed a confined electrochemical cell, and cyclic voltammograms and linear sweep voltammograms (LSVs) were recorded at predefined hopping positions. The SECCM setup is housed inside a Faraday cage with thermal isolation panels (Vaku-Isotherm) and mounted on a vibration-damping table (RS 2000, Newport) with four S-2000 stabilizers (Newport). The working-electrode potential was controlled relative to the quasi-reference counter electrode (QRCE), and the current through the QRCE was measured with a variable-gain transimpedance amplifier (DLPCA-200; FEMTO Messtechnik). An FPGA card (PCIe-7852R, National Instruments) was used for data acquisition and instrument control operated through modified Warwick Electrochemical Scanning Probe Microscopy software in LabVIEW (National Instruments). Raw data were processed in MATLAB. The electrolyte was 0.1 M $HClO_4$ with 0.1 M $LiClO_4$ as supporting electrolyte (measured pH ≈ 1.2). Measured potentials were converted to the reversible hydrogen electrode (RHE) scale using the standard conversion: ($E_{RHE} = E_{applied} + E_{Ag/AgCl(3M\ KCl)} + 0.059$ pH); with EAg/AgCl (3 M KCl) = 0.210 V.

## S2 Composition Space and Computational Details

### Composition Sampling

The Ag-Au-Pd-Pt composition space was sampled with a hybrid scheme. The lower-order boundaries (binaries and ternaries) were placed on systematic grids and the quaternary interior was filled with a Sobol low-discrepancy sequence.[2] The four pure metals were included as endpoints. Each binary edge was placed on a 1/12 grid, giving the eleven mixing ratios from 1/12 to 11/12 for each of the six element pairs, so 66 binary compositions. Each ternary face was placed on a 1/9 grid, giving the 28 interior and 27 edge points of each of the four triples, so 112 ternary compositions. The quaternary interior was filled with 400 Sobol points, drawn with `scipy.stats.qmc.Sobol` at seed 42 and mapped to the simplex by the sorted-spacings transform. Points with any element below half an atom on the 36-site slab were skipped, and points were deduplicated by their rounded percentage identity. The full set contains 582 compositions, namely 4 pure, 66 binary, 112 ternary, and 400 quaternary (Figure S2). Each pure metal was represented by a single slab. Each alloy composition was represented by three unique slabs. The boundary compositions, meaning the pure, binary, and ternary points, used rounded occupancy, in which the element counts are quantized to the nearest integer on the 36 sites and the sites are then shuffled. Because the 1/12 and 1/9 grids both divide 36, these counts are exact. The quaternary interior used per-site occupancy, in which each of the 36 sites is drawn independently from the categorical distribution set by the composition. This realizes the continuous Sobol fractions without quantizing them and adds local-environment diversity across the three variants representing each composition.

All slabs were passed through a symmetry-canonical fingerprint check during generation so that the kept variants of a composition are pairwise distinct even under symmetry. The fingerprint works in fractional coordinates and canonicalizes over the in-plane lattice translations together with the lattice-preserving point-group operations. For the hexagonal 3×3 fcc(111) cell only the identity point-group operation preserves the lattice, so the canonicalization is just over translations in this case. Pure metals collapse to a single structure. This scheme produced 1,738 distinct slabs in total, which equals four pure slabs plus 1,734 alloy slabs. From it 1,738 work function values were obtained, one per bare slab, and 15,610 hydrogen adsorption energies, which is one hollow site on each of the four pure slabs plus the nine hollow sites on each of the 1,734 alloy slabs.

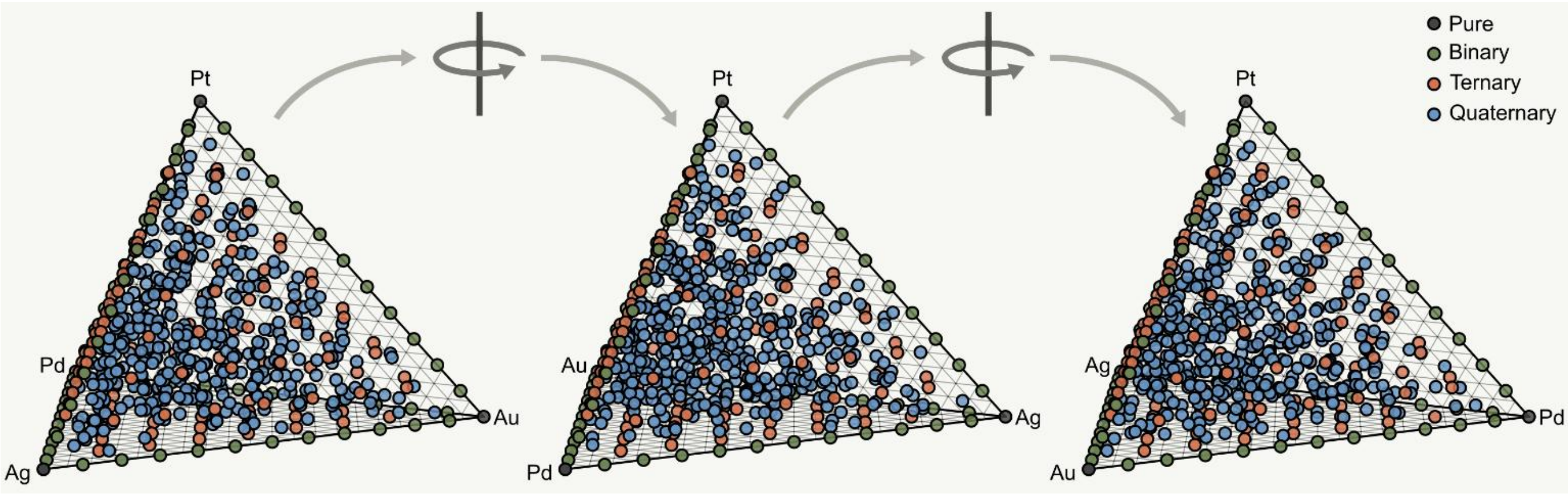

**Figure S2.** Compositions sampled for GNN training across the Ag-Au-Pd-Pt simplex space, seen from three different angles.

**Slab Set-Up and Relaxation**

Surfaces were modeled as fcc(111) 3×3×4 slabs with 10 Å of vacuum above and below. Lattice constants were assigned by Vegard's law[3, 4] as the composition-weighted average of the equation of state RPBE pure-metal fcc lattice constants: Ag 4.2087 Å, Au 4.2145 Å, Pd 3.9810 Å, and Pt 3.9942 Å. Hydrogen was placed in an fcc hollow site 1.3 Å above the surface. Each alloy slab has nine such sites, and each was occupied in a separate calculation. On the pure metals the nine sites are equivalent, so a single site was used. One hydrogen on the 3×3 cell corresponds to a fractional coverage of $\theta$ = 1/9 ML, which is the coverage of every adsorption energy reported here before the correction described below. Each configuration was first relaxed with the UMA-S-1.2 universal machine-learning interatomic potential (*via* fairchem-core version 2.20.0)[5] using the LBFGS optimizer to a maximum force of 0.05 eV/Å and the "oc20" task. In both relaxation stages the adsorbed hydrogen and the two topmost slab layers were free to move, while the two bottom layers were held fixed at their Vegard-law bulk positions. The same constraint was applied to the bare slabs used for the work function. It was then refined with RPBE[6] using GPAW (version 25.7.0).[7] The GPAW refinement used a short LBFGS relaxation to a maximum force of 0.1 eV/Å with at most ten ionic steps, which from the UMA geometry is effectively a single DFT-quality single point. The GPAW settings were plane-wave mode with a 400 eV cutoff, a 3×3×1 Monkhorst-Pack $k$-point grid, the Davidson eigensolver, Fermi-Dirac smearing of 0.1 eV, and a dipole correction along the surface normal. The self-consistency criteria were $10^{-6}$ eV on the energy and $10^{-4}$ on the density. The work function of a bare slab was evaluated as

$$\phi = V_{\mathrm{vac}} - E_{\mathrm{F}},$$

where $E_{\mathrm{F}}$ is the Kohn-Sham Fermi level and $V_{\mathrm{vac}}$ is the Hartree electrostatic potential averaged over the plane parallel to the surface and read at the cell boundary, 10 Å from the outermost atomic layer. The dipole correction removes the artificial field across the vacuum, so the plane-averaged potential is flat there. All calculations were non-spin-polarized, which is appropriate for the non-magnetic Ag, Au, Pd, and Pt. The training input is the unrelaxed ideal geometry and the label is the relaxed energy, so the surrogate is trained in the same initial-structure-to-relaxed-energy setting in which it is deployed. A molecular hydrogen energy of -6.679028 eV was used, computed at the same level of theory. The pure-metal adsorption energies and work functions for all elements presented in Figure D-F in the main text (Table S1), are computed using the exact same DFT set up as described above.

**Table S1.** RPBE-calculated values for bare-surface work function and hollow-site hydrogen adsorption energy for each of the metals presented in Figure 1D-F in the main text.

| Metal | Work function (eV) | Adsorption energy (eV, 1/9 ML) |
|---|---|---|
| Ag(111) | 4.29 | 0.39 |
| Au(111) | 5.10 | 0.32 |
| Co(111) | 4.81 | -0.41 |
| Cu(111) | 4.69 | -0.037 |
| Ir(111) | 5.37 | -0.25 |
| Ni(111) | 4.94 | -0.38 |
| Pd(111) | 5.07 | -0.46 |
| Pt(111) | 5.50 | -0.32 |
| Rh(111) | 4.97 | -0.33 |
| Mo(110)[a] | 4.42 | -0.61 |

[a] bcc.

### Coverage Correction

The repulsion was obtained by filling a surface one hydrogen at a time and tracking how the cost of each added hydrogen grows with coverage. The four pure metals were used, as a check against the literature per-metal repulsion, and the equimolar Ag-Au-Pd-Pt alloy, which carries the repulsion of interest and was sampled using ten unique slabs. All configurations were screened with the UMA-S-1.2 universal machine-learning interatomic potential to a maximum force of 0.05 eV/Å. We seek the lowest-energy filling path from the clean surface to the full monolayer. A greedy fill that keeps only the single lowest configuration at each coverage is not reliable, because a step that is lowest at one coverage can lock the path into a basin whose higher coverages are no longer the global minimum. Østergaard, Abild-Pedersen, and Rossmeisl[8] note the same limitation for their scheme and write that it gives the steepest descent from each configuration to the next but not the lowest-energy destination. We therefore use a $k$-best search of width $k$ = 5. At each coverage the $k$ lowest configurations are kept, all of them are expanded by one hydrogen, the candidates are relaxed, and the $k$ lowest are kept again. The lowest-sum nested path retained by the search is then read off at the end. On the equimolar alloy the $k$ = 5 search reproduces the exhaustive enumeration of all $2^9 - 1 = 511$ configurations and matches the per-coverage global minimum, while the greedy path lies above it and gives a jagged and biased slope. The search was run in both directions. The adsorption path fills from the clean surface and the desorption path empties from the full monolayer. The lowest-sum nested path is symmetric under reversal, so the two directions coincide once the search has converged, and the small residual difference between them is used as a convergence check. The hydrogen-slab configurations on the adsorption path and the bare slab were then refined with RPBE in GPAW at the same settings as for the GNN dataset.

Along the path we follow the differential adsorption energy for adding the $n$th hydrogen,

$$\Delta E_{\text{diff}}(n) = E(n\text{H}^*) - E\big((n-1)\text{H}\,*\big) - \frac{1}{2}E(\text{H}_2).$$

Its slope against fractional coverage is the total isotherm slope, which we denote $\alpha$. This slope mixes two effects. It contains the growing lateral repulsion and it contains the ligand effect, because the path fills strong sites first and weak sites last, so the spread of site binding energies enters the slope directly. To isolate repulsion we subtract the zero-coverage binding of the site $q_n$ that is filled at step $n$,

$$\Delta E_{\text{site}}(q_n) = E(\text{H}^*) - E(slab) - \frac{1}{2}E(\text{H}_2),$$

which leaves the residual

$$\Delta E_{\text{rep}}(n) = \Delta E_{\text{diff}}(n) - \Delta E_{\text{site}}(q_n).$$

The hydrogen reference cancels in the repulsion term $\Delta E_{\text{rep}}$. The slope of it against coverage is the mean-field repulsion, which we denote $\omega$. The subtraction removes the ligand contribution of the added site and leaves only the extra cost that comes from the hydrogens already on the surface. This differs from the centering of Østergaard *et al.*,[8] where a single global slope taken from the mean isotherm is subtracted from every site, and from Patel *et al.*,[9] where a single literature value calculated for Pt(111) is applied to every metal. Because the surrogate already predicts the site-specific dilute binding, the H-H repulsion coefficient $\omega$ is the correct quantity to layer on top of it, whereas $\alpha$, the total isotherm slope, would double-count the ligand effect that the surrogate has already learned. On a pure metal all sites are equivalent, so $\Delta E_{\text{site}}$ is constant and $\omega$ equals $\alpha$.

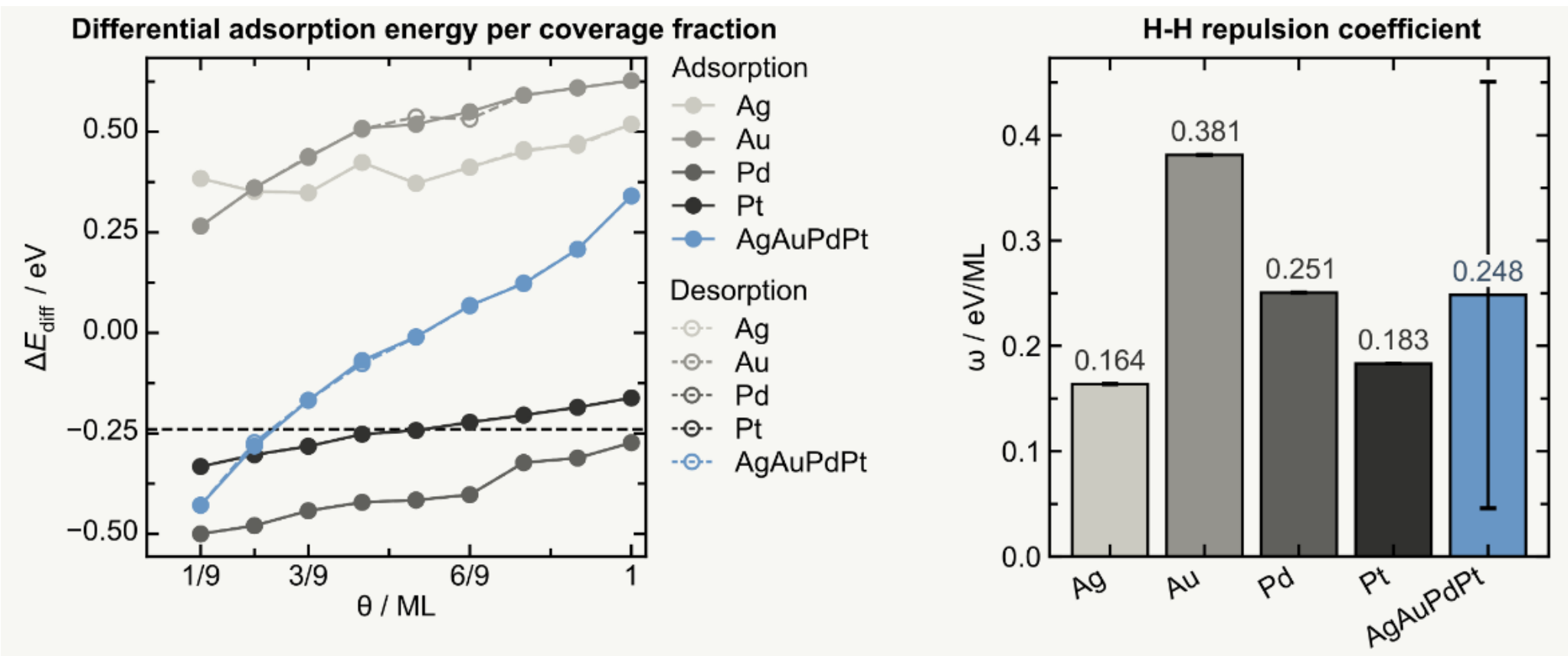

**Figure S3.** Differential adsorption energy from 1/9 to 1 ML hydrogen coverage across the four pure metals and the equimolar composition, as well as the per composition H-H repulsion coefficient $\omega$. Error bars show the standard deviation of $\omega$ across the ten distinct equimolar slabs, they represent the slab-to-slab spread rather than the uncertainty in the mean.

The equimolar mean-field repulsion is $\omega = 0.248 \pm 0.203$ eV/ML (Figure S3). The total slope is much larger at 0.789 eV/ML. The gap between them shows that most of the coverage dependence on this alloy is the ligand effect and not the repulsion. The hysteresis between the adsorption and desorption branches is within 0.02 eV/ML of zero for all compositions, which we use as a convergence check on the search rather than as a statement about hysteresis on the physical surface. The equimolar repulsion is additive. The mean of the four pure repulsions is

$$\frac{0.164 + 0.381 + 0.251 + 0.183}{4} = 0.245\,\mathrm{eV/ML},$$

a 0.003 eV/ML difference between pure metal mean and equimolar. This is expected of a genuine mean-field repulsion once the ligand effect has been removed, and it is what makes a single transferable ω meaningful. For the conversion from $\theta = 1/9$ ML to $\theta' = 1$ ML used in the main text, the rigid shift $\omega\left(1 - \frac{1}{9}\right) = 0.22$ eV, uniformly displacing the adsorption-energy distribution toward weaker binding (Figure S5).

Four approximations enter the correction as it is used in the main text. The coefficient is applied uniformly across the composition space, although the pure-metal values span 0.164 to 0.381 eV/ML and the equimolar value is additive, so a composition-weighted coefficient would in principle be available. The target coverage is fixed at 1 ML for every MA, although the working coverage is itself composition-dependent, with first-principles kinetics placing the steady-state hydrogen coverage near unity on the platinum-group metals and orders of magnitude lower on the coinage metals under HER conditions.[9] ω is taken as a single slope over the full coverage range, whereas Ref. [9] required a piecewise linear function with a coverage onset to describe the noble metals. And applying the correction as a rigid shift assumes that the shape of the site-energy distribution is unchanged between 1/9 and 1 ML, which we have not verified directly. All four are least reliable in the Ag- and Au-rich corners of the composition space.

Østergaard *et al.*[8] sampled hydrogen filling on IrPdPtRhRu(111) by evaluating every single-hydrogen addition or removal with DFT and retaining the single lowest-energy configuration at each coverage, in both directions. They defined the repulsion as the slope α of a linear fit to the nominally weighted mean differential adsorption energy against coverage, and centred the per-site distributions by subtracting that same global slope. Because the retained path fills electronically favourable sites before sterically favourable ones, α carries the ligand ordering along with the lateral repulsion, which they identify as the reason their alloy repulsion (0.35 eV/ML) exceeds their pure-metal values (0.09-0.22 eV/ML). Our scheme is theirs with the retained beam widened from one configuration to five, which on the equimolar Ag-Au-Pd-Pt slab reproduces exhaustive enumeration of all $2^9 - 1 = 511$ configurations and, because the lowest-sum nested path is reversal-symmetric by construction, closes the adsorption-desorption gap to within 0.02 eV/ML. The residual gap therefore serves as a search-convergence diagnostic rather than a physical result. We additionally subtract the zero-coverage binding of each newly filled site before taking the slope, which is not the global centering used in Ref. [8] and which isolates ω from α. On Ag-Au-Pd-Pt, whose site energies span a wider range than IrPdPtRhRu, we get α = 0.79 eV/ML and ω = 0.25 eV/ML, so the ligand share is larger still, and ω, unlike α, returns to the mean of the constituent pure metals. This quantifies the ligand contamination that Ref. [8] identified as the origin of the apparent repulsion enhancement on alloys. Our per-metal values differ from theirs for Pd (0.25 versus 0.15) and Pt (0.18 versus 0.22), which we attribute to differences in relaxation approaches, such as the relaxation of the two top layers together with the hydrogens done here versus the fully frozen slabs in Ref. [8].

## S3 Graph Neural Networks

### Graph construction and models

Periodic graphs were built from the atomic simulation environment (ASE) (version 3.28.0)[10] routine neighbor list (“ijdDS”) with a hard radial cutoff. Atomic positions were wrapped into the cell. For each atom the neighbors were kept up to a maximum of 50 per atom, retaining the shortest distances, and periodic images were encoded by integer cell offsets. The source-to-destination displacement of an edge was computed as $\mathbf{r}_{ij} = \mathbf{r}_j + \mathbf{S}_{ij}\mathbf{h} - \mathbf{r}_i$. Here $\mathbf{r}_{ij}$ is the displacement from source atom $i$ to the destination atom $j$, $\mathbf{r}_i$ and $\mathbf{r}_j$ are the Cartesian positions of the atoms wrapped into the cell, $\mathbf{S}_{ij}$ is the integer cell-offset identifying the periodic image of $j$ that the edge connects to, and $h$ is the cell matrix whose rows are the lattice vectors, so that $\mathbf{S}_{ij}\mathbf{h}$ is the lattice translation to that image.

Two graph neural network (GNN) surrogate models were trained, one for the hydrogen adsorption energy and one for the work function. Both deployed models are DimeNet++,[11, 12] built from the DimeNetPlusPlus implementation in PyTorch Geometric (version 2.6.1).[13, 14] They use 128 hidden channels, three interaction blocks, an 8 Å cutoff, six radial basis functions, seven spherical basis functions, a polynomial envelope with exponent five, and sum pooling. The single scalar output is the target value in eV. The models are trained in the initial-structure-to- relaxed-energy setting, with the unrelaxed ideal geometry as input and the relaxed-structure value as label. As a reference for the GNN performance, a parameter-free pure-metal linear-combination baseline was used, in which the prediction for a given composition is the composition-weighted sum of the pure-metal DFT values,

$$\hat{y} = \sum_{m \in M} x_m \, y_m, \qquad M = \{\mathrm{Ag}, \mathrm{Au}, \mathrm{Pd}, \mathrm{Pt}\},$$

where $x_m$ is the atomic fraction of element $m$ in the given slab, and $y_m$ is the pure-metal value of the target $y$, with prediction $\hat{y}$.

**Training Protocol**

Targets were used in raw units without standardization. The training loss was the L1 loss, which equals the mean absolute error (MAE) in eV. For hydrogen adsorption, datapoints with $|\Delta E_{\mathrm{H}^*}| > 2.5$ eV were filtered out. For the work function, slabs with $\Phi < 4$ eV were dropped. The networks were trained with AdamW[15] at a weight decay of 0.01 and gradient clipping to a maximum norm of 10. The initial learning rate was $5\times10^{-4}$, reduced on a validation-MAE plateau by a factor of 0.5 with a patience of ten epochs and a minimum of $10^{-6}$. The batch size was 32. The maximum number of epochs was 1000, with early stopping after 100 epochs without an improvement in the validation MAE. The saved model was the checkpoint with the lowest validation MAE. The data was split 80/10/10 into train, validation, and test. The split used a fixed seed of 42, the grouping unit was the composition, so all variants and all hydrogen sites of a given composition were kept in the same partition. This was done to avoid leakage across the split. To check that the held-out test compositions are representative rather than clustered, the mean nearest-neighbor distance between test compositions in element-fraction space, 0.129, was compared with a type-stratified random baseline, 0.126 ± 0.006, placing the actual split at the 66th percentile. The surrogate errors are therefore interpolation errors within the sampled alloy space, not extrapolation errors to unseen alloy families.

**Evaluation**

The deployed adsorption-energy model reached a test MAE of 0.0093 eV, a test root mean square error (RMSE) of 0.0198 eV, and $R^2$ = 0.994 on the 1,567 held-out adsorption energies. The deployed work-function model reached a test MAE of 0.0179 eV, a test RMSE of 0.0278 eV, and $R^2$ = 0.988 on the 175 held-out slabs. These numbers measure the surrogate accuracy relative to the RPBE data within the sampled alloy space. They do not include systematic uncertainty from the exchange-correlation functional or the slab set-up. The pure-metal linear combination baseline gave a test MAE of 0.1646 eV and $R^2$ = 0.361 for the adsorption energy, confirming that the adsorption energy is primarily a local quantity. For the work function the same baseline gave a test MAE of 0.0997 eV and $R^2$ = 0.771, showing that the work function carries a strong composition trend (Figure S4).

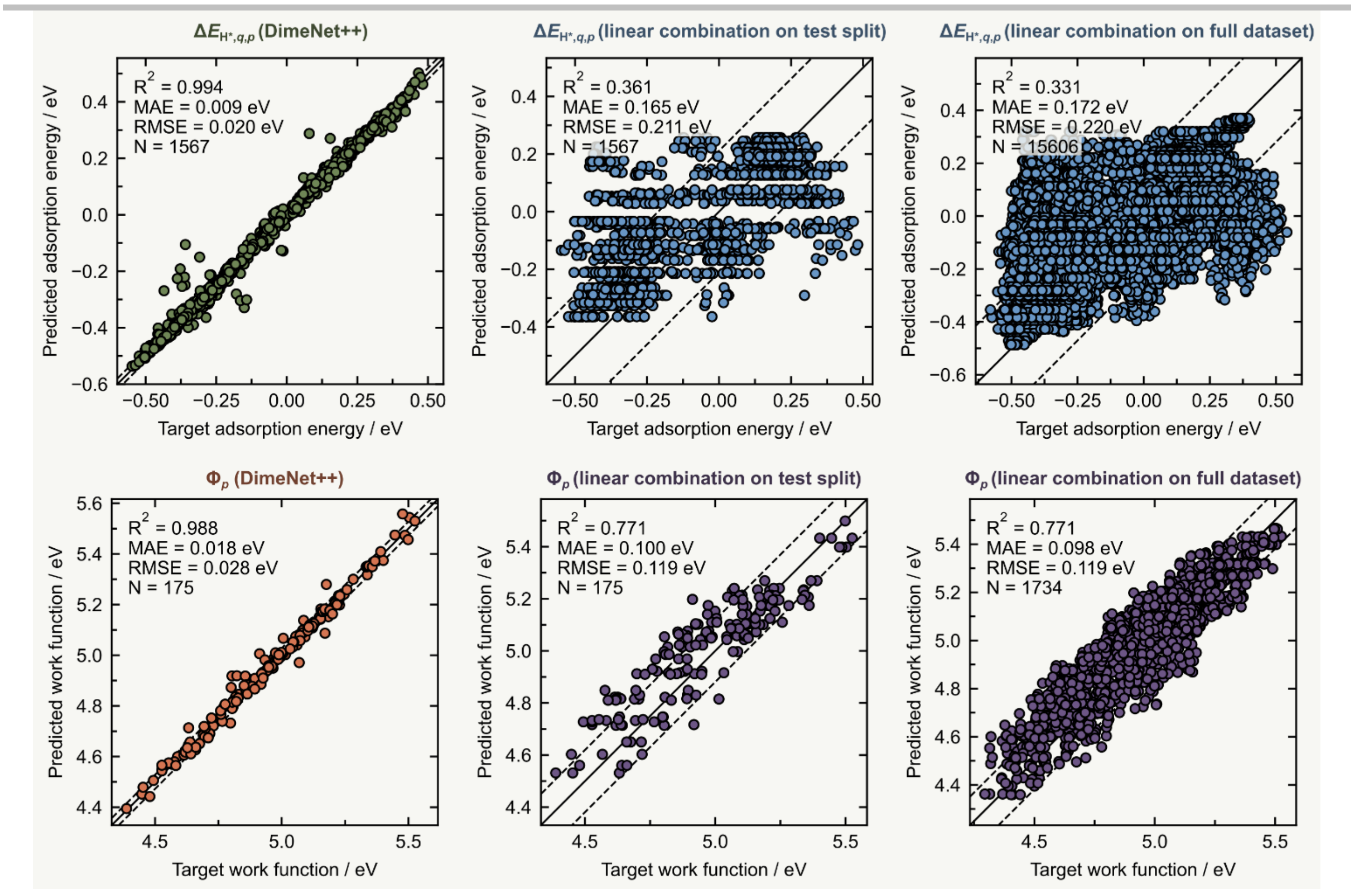

**Figure S4.** Evaluation of the two GNN models and their respective linear combination baselines. Top row is adsorption energy, comparing the DimeNet++ predictions on the test split versus a linear combination of the pure metal values on both the test split and on the complete dataset. The bottom row is the same but for work function.

## S4 Descriptor Inference

The trained models were applied to every EDX composition, for each measurement area (MA), resulting in adsorption energy distributions and mean work functions per MA. Since the atomic arrangement within a MA is unknown, each MA was represented by an ensemble of random fcc(111) 3×3×4 slabs generated from the normalized EDX composition. The slab geometry and lattice constants matched the training slabs, with the lattice assigned by Vegard's law from the MA composition. Sampling used the per-site occupancy mode, in which each of the 36 atoms is drawn independently from the categorical distribution defined by the elemental fractions. Individual slabs therefore carry multinomial fluctuations around the EDX composition while the ensemble average recovers it. All four layers were drawn from the same distribution, so surface segregation and layer-dependent enrichment are not included. The coverage correction of Section S2 is likewise applied identically to every MA, so it is a library-wide approximation rather than a per-MA quantity. For each MA, $N$ = 1000 random slabs were generated. Hydrogen was placed at each of the nine fcc hollow sites of every slab, giving N × 9 = 9000 predicted adsorption energies per MA, and the same 1000 bare slabs gave N = 1000 work function values, which were averaged to obtain the MA work function $\Phi_p$. Since every variant of a slab representing a composition shares the same Vegard-law lattice, the periodic neighbor-list graph was built once per distinct geometry, that is, once per adsorption site and once per work function, and reused across all variants by swapping only the atomic-number node feature. This graph caching is bit-exact with rebuilding every graph and reduced the inference wall-clock time by about a factor of six on the GPU, which made the roughly nine million adsorption-energy evaluations across all MAs of the three libraries more manageable. Figure S5 shows the resulting adsorption-energy distributions for the four composition extremes of each library, at both 1/9 and 1 ML.

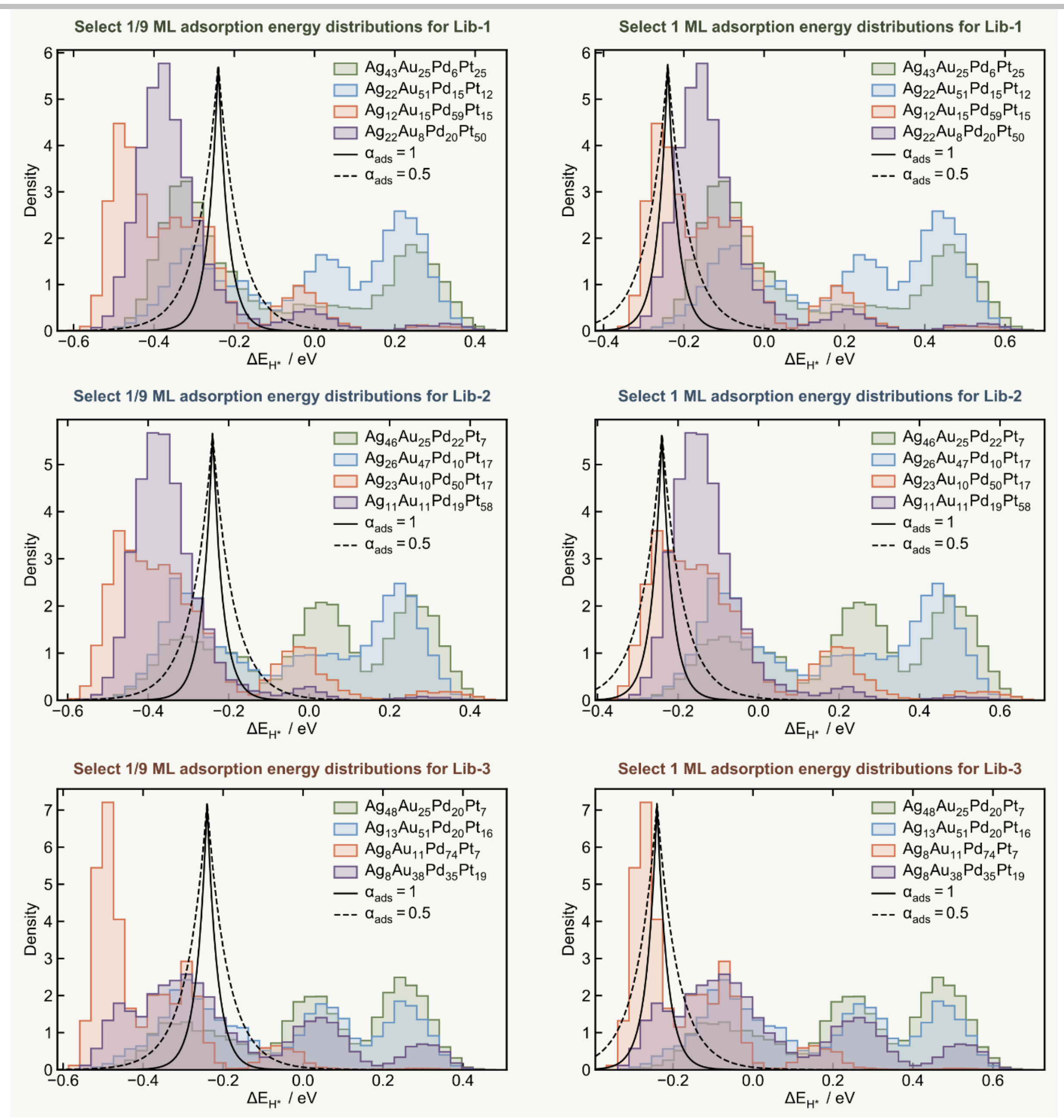

**Figure S5.** Inferred adsorption-energy distributions for the four composition extremes of each materials library, that is, the compositions with the highest atomic percentage of Ag, Au, Pd, and Pt. Left column, dilute 1/9 ML distributions and right column, coverage-corrected 1 ML distributions. Rows are Lib-1, Lib-2, and Lib-3.

## S5 Fitting Protocol

### Loss, Mask, and Optimizer

All fits minimized a Huber loss[16] on the log-current residuals. The residual for a cell is

$$r = \log_{10}[\max(-j_{\text{model}}, j_{\text{floor}})] - \log_{10}[\max(-j_{\text{exp}}, j_{\text{floor}})],$$

with the current floor $j_{\text{floor}}$ = $10^{-9}$ A/cm$^2$, and the Huber transition at 0.3 decades. A cell is one MA at one potential, and a cell entered the fit only if the experimental current was cathodic and above the floor, meaning $j_{\text{exp}} < -10^{-9}$ A/cm$^2$. This filter was applied per potential, so a MA could contribute at one potential and be excluded at another. The same valid-cell mask was used for fitting and for scoring. The four analysis potentials were -100, -200, -300, and -400 mV versus RHE, and the primary fit is the joint fit over all four potentials. At a single potential the Tafel factor and $j_0$ are not separable, so single-potential fits were used only as diagnostics. The objective was minimized with `scipy.optimize.least_squares` using the trust-region reflective algorithm and an exact Jacobian from automatic differentiation, evaluated in double precision. The exponential arguments were clipped to [-700, 700] and the sigmoid arguments to [-50, 50] before evaluation. All parameters except $\alpha_{\mathrm{wf}}$ were optimized in an unconstrained transformed space. The current scale and the transport limit were parameterized as $j_0 = \exp(p)$ and $j_{\mathrm{lim}} = \exp(p)$, enforcing positivity. The transfer coefficient and the volcano width were parameterized as $\alpha_{\mathrm{c}} = \mathrm{sigmoid}(p)$ and $\alpha_{\mathrm{ads}} = \mathrm{sigmoid}(p)$, enforcing values between 0 and 1. The work-function coupling was optimized directly with the explicit box $-1 \leq \alpha_{\mathrm{wf}} \leq 1$. Each fit used several starting points and the solution with the lowest optimizer cost was kept. The starting current scale and transport limit were set from the mean and maximum of the valid experimental currents. Goodness of fit is reported as the coefficient of determination in log-current space,

$$R^2_{\log} = 1 - \frac{\sum_i \left(\log_{10}|j_{\mathrm{exp},i}| - \log_{10}|j_{\mathrm{model},i}|\right)^2}{\sum_i \left(\log_{10}|j_{\mathrm{exp},i}| - \mathrm{mean}\left[\log_{10}|j_{\mathrm{exp}}|\right]\right)^2},$$

evaluated over the same valid cells used for fitting. Lib-1 and Lib-2 hold 342 MAs and Lib-3 holds 341. With the cathodic-and-above-floor mask, the number of valid multi-potential cells was 1368, 1318, and 1364 for Lib-1, Lib-2, and Lib-3.

### Within-Library and Transfer Fitting

Within-library fits were performed independently for each library, with all parameters library-specific. Between-library transfer used a leave-one-library-out scheme. The shared descriptor parameters, meaning $\alpha_{\mathrm{c}}$, $\alpha_{\mathrm{ads}}$, and $\alpha_{\mathrm{wf}}$, were fitted jointly on two libraries and frozen. Three levels of transfer were then compared on the held-out library. Direct transfer, where nothing is recalibrated on the held-out library. Scale-adapted transfer, where the current scale $j_0$ is recalibrated. Scale-and-transport adapted transfer, where both $j_0$ and the current limit $j_{\mathrm{lim}}$ are refit on the held-out library. The recalibration used repeated MA-wise cross-fitting, in which all four potentials of a MA are held out together, the nuisance parameters ($j_0$ and $j_{\mathrm{lim}}$) are fitted on the remaining MAs, and the held-out MAs are predicted without further adjustment. The cross-fitting used five folds and 50 repeats. The reported transfer standard deviations are the sample standard deviations over the repeated cross-fitting folds. The per-library difference between within-library and leave-one-library-out fit quality is shown in Figure S6 for each model and adaptation level.

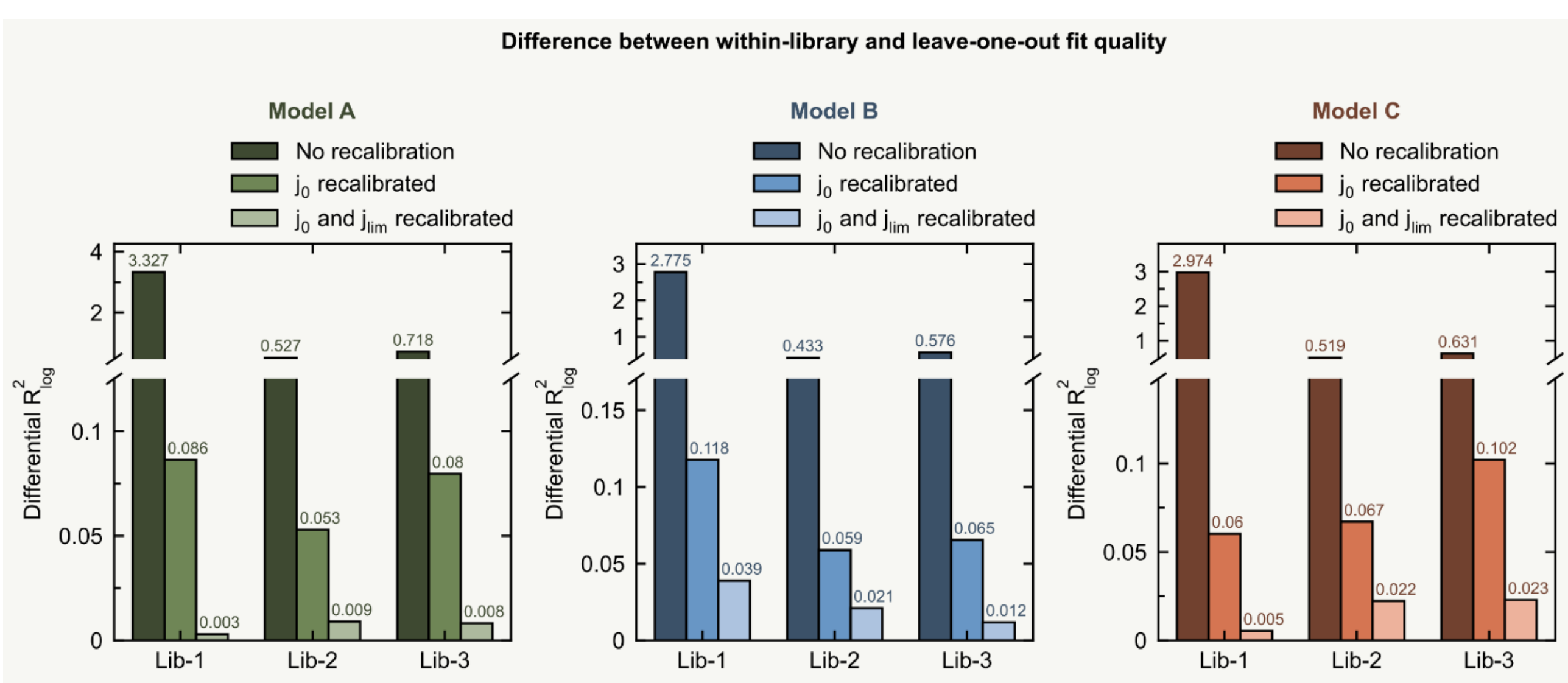

**Figure S6.** Difference in $R^2_{\log}$ between the within-library fit and the leave-one-library-out fit, for each model and each level of recalibration (no recalibration, recalibration of $j_0$ only, and recalibration of both $j_0$ and $j_{\mathrm{lim}}$.)). Note the split vertical scale.

### Data availability

The data that support the findings of this study are openly available in Zenodo at https://doi.org/10.5281/zenodo.22031607.